# TWEETING FOR THE CAUSE:
## Network analysis of UK petition sharing[1]

Peter Cihon[1], Taha Yasseri[2], Scott Hale[2], and Helen Margetts[2]

[1] Williams College, Williamstown, MA 01267, USA
[2] Oxford Internet Institute, University of Oxford, Oxford OX13JS, UK

**Abstract**
Online government petitions represent a new data-rich mode of political participation. This work examines the thus far understudied dynamics of sharing petitions on social media in order to garner signatures and, ultimately, a government response. Using 20 months of Twitter data comprising over 1 million tweets linking to a petition, we perform analyses of networks constructed of petitions and supporters on Twitter, revealing implicit social dynamics therein. We find that Twitter users do not exclusively share petitions on one issue nor do they share exclusively popular petitions. Among the over 240,000 Twitter users, we find latent support groups, with the most central users primarily being politically active 'average' individuals. Twitter as a platform for sharing government petitions, thus, appears to hold potential to foster the creation of and coordination among a new form of latent support interest groups online.

## I. Introduction:

Internet-based platforms facilitate citizen participation in government, and government-sponsored petition websites—available in a number of democracies, including the United Kingdom and the United States—are an increasingly popular example. On such platforms, citizens undertake a 'tiny act' of political participation (Margetts et al., 2015), expressing their policy preferences to government and lobbying for change. Petitions have a rich history in the UK, and online petitions, in some ways, present a continuation of that tradition (Bochel 2012). Although petitions are not new, online petition platforms—together with social media—offer a means of circulating and tallying signatures on a scale never seen before. This paper analyzes the relationship between sharing activity on social media and petition signature outcomes. This analysis uses approximately 1 million tweets over a 20 month time span during which more than 11 thousand petitions from the UK government petition website were shared.

---

[1] **Acknowledgment**
The data collection and preliminary analysis for this paper was carried out as part of the ESRC professorial fellowship, 'The Internet, Political Science and Public Policy: Re-examining Collective Action, Governance and Citizen-Government Interactions in the Digital Era' RES-051-27-0331 (2011-2015), and we thank the ESRC for their support. We would also like to thank Alex Paseltiner, Reid Pryzant, Matt McNaughton, Matt Baya, and Williams College OIT for computational assistance during this project.

Quantitative analysis of online petition campaigns is a recent focus in literature on collective action and political participation. Twitter and other social media platforms offer researchers 'trace data' of individuals' interactions, permitting large-scale analysis for the first time. Researchers have widely studied these trace data in varied contexts across the world, particularly focusing on collective action surrounding political events (e.g., Gonzalez Bailon et al., 2011; Cihon and Yasseri 2016; Margetts et al., 2015). Among social media platforms, Twitter provides one way to study how people share petitions. The Twitter platform facilitates the spread of particular messages through re-tweets, which can cascade through many individual users' networks (e.g., Goel et al., 2016; Bakshy et al., 2011; Gonzalez-Bailon 2011). Use of hashtags links conversations across individual networks, and popular hashtags are further publicized by the platform itself. Moreover, Twitter provides researchers with data on public tweets through its API.[2] This work makes use of this Twitter data in conjunction with UK Petition website outcomes.

The UK government established its first petition platform, the No. 10 Downing Street petition website, in 2006, and it received some eight million signatures from five million unique email addresses between its founding and 2010.[3] In 2010, the No. 10 Downing Street website was replaced by a second platform run by the Coalition Government from 2011 to 2015.[4] This second platform closed with the dissolution of parliament ahead of the 2015 general election, and was subsequently replaced by the current, third instantiation of the platform.

[2] Application programming interface, a means for researchers to obtain data from Twitter.
[3] https://web.archive.org/web/20120113080700/http://www.mysociety.org/projects/no10-petitions-website/
[4] https://web.archive.org/web/20150330012917/http://epetitions.direct.gov.uk/

Structure and procedure for petition websites vary, though the broad concept is as follows. Individuals may create a petition on a particular topic of concern and then make the petition viewable for others to sign. If the petition garners sufficient signatures, then it will receive a government response. In the case of the Coalition Government platform we studied, only British citizens or residents were supposed to write and sign petitions. Petitions with 10,000 signatures received a government response and those with 100,000 or more signatures were considered for debate in Parliament.[5]

Since their launch, government petition platforms have been the subject of a number of investigations. Dumas et al. (2015) examined a small sample of petitions' signature clusters on the US Government We the People website. The authors offered largely descriptive accounts of participation themes within the platform and, particularly, gun-control. But these findings came from the simplistic filtering of data to exclude all but the most active users. We draw inspiration from these thematic analyses, and use more developed statistical filtering methods that permit more nuanced findings. Hale et al. (2013) analyzed signature growth of over 8,000 petitions submitted to the first UK website over a two year period, and found that only 6 percent of petitions received enough signatures to cross the then-response threshold of 500 signatures, and of those, 43 percent crossed the threshold on the same day they were posted. Yasseri, et al. (2013) reproduced these analyses on the UK Coalition Government website (the same one as studied in this paper), and found similar results. The authors argued that social media may encourage the rapid lifecycles of petitions. The noted importance of early signatures brings into question the methods of publicizing petitions. Margetts, et al. (2015) offer analysis of petition platforms and social media publicity, including a time-shift analysis of tweets and signatures that indicated that tweets drove signatures, not the opposite. This work goes further by using network analysis to understand the relationship between social media and petition signatures.

[5]https://web.archive.org/web/20150326050709/http://epetitions.direct.gov.uk/how-it-works

Signatures may or may not cross a response threshold, but the meaning of “success” varies. Indeed, even for petitions that cross a threshold, the government response may not be what the petition author desires. Wright (2015) offers a qualitative study of petition outcomes for the first UK petition platform. Semi-structured interviews reveal that those whose petitions received official replies were commonly unhappy with those replies. Often the replies were similar in tone and substance to previous communications on the subject. Moreover, 300 petitions did not receive a reply despite crossing the threshold, and many more received a reply only after an extended delay.

Dumas et al. (2015) offer a theoretical framework that places petition platforms within a broader political context. Petitions may offer an instantiation of agenda setting within punctuated equilibrium theory. In this theory, events disturb policy equilibria and create moments in which policy change may occur. Viewed in this way, petitions may be one channel through which events may lead to such change. On this note, Bochel (2012) analyses five case studies of government procedures surrounding petitions. She finds that the second UK petition website was more “descriptive” than “substantive,” meaning individual petitions’ content was not scrutinized and policy actions were rarely taken based on such petitions. Systems that do not offer substantive participation for authors of petitions risk demoralizing citizens and undermining democratic legitimacy. Although the current work does not interrogate these questions, they remain important considerations for interpreting our findings. Indeed, we find that, in correspondence with agenda setting within punctuated equilibrium theory, petitions often respond to particular events and, as one might expect, seek to affect policy.

Whether or not they are achieved, petitioners' goals pertain to policy. Individuals may use petitions to address their wishes to government or to seek redress in case of government misdeeds (Lidner and Riehm 2011; Bochel 2012). Petitions may empower individuals or they may be used by traditional interest groups to stir up support for their own goals. Or they may help form new advocacy coalitions and foster coordination between activists, as described by Strange (2011) in his analysis of Global Group Petitions. Our work helps inform these questions with an understanding of precisely who the Twitter sharers of widely shared and widely signed petitions are. Lidner and Riehm (2011) surveyed groups of Germans who submitted petitions offline and online in Germany. The authors found that although authors of online petitions were younger than those of offline petitions, online petitioning appeared to "amplify existing inequalities in participation" that supported disproportionately wealthy and politically active individuals. Focusing on petition-signing behavior, Margetts et al. (2015), which argued that by extending the 'ladder' of political participation, petitions and other forms of online collective action could bring new groups into the political process. Although Twitter data is not well suited to answer such questions, we will offer some conjecture based on self-provided information in Twitter bios.

## III. Data

This work uses Twitter data that was collected that spans approximately 20 months, from July 2013 to March 2015 using the Twitter Search API as part of research conducted for the ESRC professorial fellowship "The Internet, Political Science and Public Policy." The Search API is preferable to the Streaming API in order to be able to capture tweets with links to petitions using link shortening services. We did not encounter any rate limiting issues the number of tweets linking to a petition on the UK petition website is, in general, a small percentage of all Twitter activity. Given the extended time period of the data collection it was inevitable that we would experience a small amount of lost data due to network connectivity and electricity issues; however, this should not produce any systematic biases as the timing of interruptions was more or less random. During the period of data collection, over 1 million tweets linking to 11,706 petitions were collected. Each tweet is associated with considerable information about the post and its posting user. From a dataset of all tweet data, we extracted the following fields for further analysis: tweet content, time posted, favorites, retweets, and authoring user's information. The latter includes all profile information, including images and bio, total number of tweets, number of followers, number of users the account is in turn following, whether or not the account is verified, and the date the account was initially created. This data offers rich insight into the content of tweets, their reception within users' networks, as well as the identities of posting users. These data are then linked to petition information—title, date posted, and number of signatures—obtained from the UK petition website at the time.

## IV. Methods

Here we briefly discuss our analytical methods. Preliminary statistical analyses follow standard Ordinary Least Squared regression model methodologies. This paper considers the projections of a two-mode network. Whereas one-mode network analysis may analyze micro-level or macro-level relationships, it does so separately. In the present study, one-mode data would permit analyses of individual Twitter users' relationships with one another or individual petitions' statistical relationships with other petitions on the UK petition website. Such insights are useful, but they do not reveal the relationships that tie individual Twitter users to each other *through* petitions. Two-mode data, however, permit such an analysis of both micro- and macro-level relationships. Our analysis uses two projections of the two-mode network. In the first, individual Twitter users are connected if they both have tweeted about the same petition, regardless of their Twitter following/follower relationship. In the second, petitions are connected if the same user has tweeted about both of them.

In both projections, these connections, or edges, are undirected and weighted. In this way, relationships are not one-sided but reciprocal. Edge weights warrant further description. Given the large size of our dataset, many observed connections could simply be statistical noise indistinguishable from random chance. To account for this, we alter edge weights to reflect a measure of significance above random chance. Drawing inspiration from statistics of word occurrence in natural language (Yan and Yasseri, 2016) and methods to check edge significance in communication networks (Gillani et al., 2014), we set the edge weight according to:

$$weight = \frac{log_2 \frac{P(x,y)}{P(x) * P(y)}}{-log_2[P(x,y)]}$$

Where *P(x,y)* is the probability of connections between nodes $x$ and $y$, given all connections extant in the network, and *P(x)* and *P(y)* represent the probability of nodes $x$ and $y$ having the as many connections as they do. This is then adjusted with logarithms and connection probability in the denominator. The result yields a *normalized* ratio of the connections between two nodes relative to all of those nodes' connections, and this replaces the weight for a given edge in the network. Thus, if there is a large number of connections between two nodes *relative to* all of those nodes' connections, that particular relationship has a high significance and is represented with a large edge weight value.

Due to the large size of the networks analyzed, we utilize filtering methods to facilitate analysis with available computational systems. This is practically necessary and is standard practice in the field (See Cihon and Yasseri, 2016). Upon generating the petition network projection, some 11,000 petitions were connected by more than 40 million edges with each edge connecting two petitions linked to by at least one common Twitter user. The unfiltered user projection connects some 250,000 users with 977 million common petition relationships. Although there exist many ways to filter networks, whether by random subsample or particular users of interest, we elect to filter based on edge weight. In so doing, we filter for the most statistically significant relationships for our analysis, permitting us to be confident that the observed connections did not occur by random chance. We filter the petition network such that only the 10 percent most significant edges are addressed in further analyses, and we filter all but the top 5 percent for the user network.[6] In our petition projection, we use further filter to remove lone petitions that were not tweeted by common users with significant weights yielding a network of 9,090 petitions with 3.98 million edges. The filtered user projection contains 241,506 users connected by 48.9 million edges. Unless otherwise noted, all analyses use these two filtered networks.

In order to analyze these networks, first we seek to look for similarities in petitions based on their connections via common Twitter users. This method uses network structure, i.e., connections between nodes and their weights, to discern underlying patterns in qualitative theme of the petitions. To this end, we use the Louvain modularity method of community detection, which efficiently assigns nodes (petitions or users) to clusters when their connections are above random chance of connection (Blondel et al., 2008). This method is the standard in large network community detection.

[6] These filter thresholds were selected to meet limitation in computational power.

We subsequently analyze the centrality of nodes within the network. As edges in our networks are a function of edge significance, central nodes are determined by frequency of significant relationships as well as a diminished effect for that same metric of its connections. That is, a central nodes connected to a less central petition will raise the centrality of the latter somewhat. This method, adapted to undirected graphs, is conducted using the PageRank algorithm (Page et al., 1999).

## V. Analysis

We first offer preliminary analyses of the linear relationships between number of tweets, users, and signatures for petitions. We then use a petition network to unpack what types of petitions are shared on Twitter. We next describe tweeting behavior of these petitions. We lastly analyze the most active users and their implicit interest groups. An integrative discussion of all results subsequently follows.

### A. Correlations and OLS Regression Analyses

We begin with some preliminary correlation and regression analyses. Each petition has a number of signatures, tweets, and Twitter users associated with it. The distribution of these associations is quite skewed (See Table 1). For this reason, we use the natural logarithm of all variables for correlation and regression analyses. We find a high correlation between signatures and tweets (0.70), signatures and unique users tweeting (0.72). Notable here, even with logarithmic variables, the distribution of signatures and tweets across petitions remains quite skewed, with a few petitions receiving a large share of signatures/tweets (See Figures 2 and 3).

Next we run a series of Ordinary Least Squared regressions (See Table 2). For a given increase in tweets, there is a disproportionately large increase in signatures: Regression 1 shows that for a 10 percent increase in the number of tweets about a given petition, we expect an 11.4 percent increase in petition signatures. This, together with the time-shift analysis on similar data in Margetts, et al. (2015), indicates that petitions receiving more attention receive even more signatures, raising the importance of publicity via tweets. More than simply publicity, however, an increase in unique users who are tweeting is associated with an *even* more disproportionately large increase in signatures; as shown in Regression 2, holding number of tweets constant, a 10 percent increase in the number of unique users mentioning a petition yields a 14.9 percent increase in signatures. This means that the number of users talking about a petition on Twitter is even more strongly associated with the number who signs it. This is intuitive as only British citizens or UK residents with a valid email address can sign a petition, each *at most once*. This would lead one to expect users to be more closely associated with signatures than simply the number of petition-related tweets they may tweet. But what if an individual signs the petition after seeing a tweet but does not think to tweet it again herself? Such an individual would escape analysis in Regression 2. In Regression 3, we generate a value for the audience, the total number of Twitter users who saw links to each petition.[7] Audience is a less strong—yet still significant—predictor: a 10 percent increase in unique audience yields a 4.8 percent increase in signatures. Regression 4 examines maximum possible exposure of all tweets on the hypothesis that an individual within a network seeing two tweets about the same petition from a user may be *more* likely to sign it. This proves to not be the case, as a 10 percent increase in total audience yields only a 4.6 percent increase in signatures, less than the unique audience described in Regression 3.

[7] The audience is derived from the size of individual users' follower counts, summed. Individual users may be over-counted, as a given user could follow multiple users in the sample. The individuals in the audience are not necessarily unique, but they will see a link to a given petition only once, in contrast to maximum possible exposure, used below.

Regression 5 predicts signatures by including tweets, users, and, as a proxy of petition topic, a series of variables for the government department to which the petition was addressed. Including all of these variables in the model yields the highest R-squared of all the regressions, but it is only 0.536, meaning that more than 46 percent of the variation in signatures remains unaccounted. We turn to other analyses to explain this yet-unexplained variation.[8]

**B. Petitions**

Significant yet overlooked in the regression analysis is the ways in which people share petitions. We use the petition network connected by common users to analyze what characteristics lead users to share petitions and what effect those characteristics and resulting sharing behavior have on signature outcomes. In the network, 9090 petitions are connected by 3.98 million edges, each edge having a weight proportional to the number of users who tweeted about both connected petitions. Of particular interest, 178 of 11,706 petitions crossed the 10,000-signature threshold to receive a government response. Figure 3 shows a word cloud of these petitions' titles, with words log-scaled by the number of occurrences. Notably, these petitions appear to address a variety of topics. This raises questions as to the relationships between petitions: are users signing the same petitions? Twitter data holds the answer.

This analysis focuses on Twitter sharing activity-links between petitions. For our analyses we divide this interest into three questions accessible with distinct methodologies:

[8] It is important to note that although all coefficients on the above regressions have been statistically significant by traditional thresholds (See Table 2), we make no claims about statistical power nor do we claim any causal effects. With such a large dataset, statistical significance as measured by p-values means little (See Vidgen and Yasseri 2016). We make no causal claims about Twitter sharing and petition signatures from these regressions because they are largely uncontrolled; instead, we use them to explore statistical relationships that motivate our further network analyses.

(1) are petitions that have similar topics more likely to be shared by the same users? The answer to this question determines if users are issue-oriented or they share petitions with a diverse set of topics.

(2) are more successful petitions more likely to be shared by the same user than those with fewer signatures? Here, observation of an assortative network, i.e., high signature correlation between petitions campaigned by the same users, indicates that the related petitions do not compete with each other, but rather success comes in clusters.

(3) are petitions with many signatures similarly important in tweets, i.e., are petitions outcomes related to tweets about petitions? The answer to this question sheds light on the role of social media in political campaigns.

The first question implicates community detection. We run the Louvain Modularity community detection algorithm on the petition network, yet find the yielded clusters of very low quality (modularity-quality score 0.025 / 1.000). Given that this data spans two years, it is possible that such petition topic clusters may be unstable over time, with a single user tweeting about animal rights two years ago but about welfare reform today, as her interests change. If this were the case, we might expect that short-term networks would have clear petition topic clusters like our 2 month preliminary analysis. In this light, we break the network apart into separate time-based networks.[9] We then run the Louvain Modularity community detection algorithm on each, but find similar results: there are not coherent subgroups of petitions that are co-shared. These results and their importance will be analyzed in the following section.

[9] One year intervals (before 2014, 2014, 2015) were chosen based on data limitations. Unlike the original 2 month sample, which drew on tweets from two months but petitions from much earlier, we generated these sub-networks using petition dates—not Twitter dates. As such, we sought to incorporate the wider range of petition origins that was present in the preliminary study as well.

Our second motivating question, are similarly successful petitions more likely to be shared by the same Twitter users, requires different analyses. Here we compare the weight of edges between petitions with a measure of the connected petitions' signatures. Recall that edge weight reflects the number and significance of common users between two petitions. The measure of petition signatures is the signatures of each multiplied together.[10] One might expect users to simply share high-signature petitions if these individuals are more likely to see such petitions tweeted or publicized elsewhere. The results indicate that is not, in fact, the case: the log-log correlation between signatures and significance of relationship (edge weight) is only 0.04, meaning that a relationship is almost non-existent. Thus, users who share multiple petitions often share both successful and unsuccessful petitions—a finding similar to that found by Huang et al. (2015) regarding power users on Change.org.

That said, to what degree are successful petitions important in the sharing network? This question raises the salience of centrality measures. Centrality helps discern if sharing of high-signature petitions is integral to the petition-sharing network, or if sharing activity is less important for high-signature petitions. Notably, petitions that are central in the petition network are not the most successful: the log-log correlation between PageRank scores and signatures is quite low: 0.1322. Petitions with many signatures are not necessarily shared by users who share other petitions. Yet, this is not definitive. Figure 4 plots the relationship between log PageRank and log signatures, and there is some notably variation among high-signature petitions. Indeed, when considering only petitions that have 10,000 or more signatures, the log-log correlation is much higher: 0.4915. Among the most widely signed petitions, there is a positive relationship between Twitter users sharing them along with other petitions and their signature outcomes. This relationship is quite weak more broadly, however, and this makes us question whether the widely signed petitions are simply being tweeted after crossing the threshold as opposed to helping gain signatures. This is a question that will be explored further below.

[10] This is done to account for both connected petitions' signatures in a single value for each edge, which can then be compared to the edge weight.

Moreover, it is important to note that although 178 petitions crossed the response threshold, only 110 of them appear in the filtered petition network. This means that 38 percent of successful petitions did not achieve common Twitter activity at our significance threshold.

### C. Tweets

Our dataset draws on over 1 million tweets for analysis. Tweets facilitated the creation of the petition network discussed above as well as the user network discussed in the following section, and our focus is primarily on these two networks. These networks reveal much about the most active Twitter users, but what of the general population? Here we use all of the Twitter data—unfiltered for significance unlike the networks—to provide additional insight. Recall that past studies have found that signatures within the first day of posting a petition are very important in determining its success (Hale et al. 2013; Yasseri, et al. 2013). With this in mind, we analyze the time gap between the creation of a petition and when the linking tweet is posted. The results are surprising: the length of the delay is *positively* correlated with the number of tweets (log-log correlation: 0.5219). Based on prior studies we would instead expect a highly negative correlation. The distribution of time delays (Figure 6) appears bimodal. The near-zero mode represents initial promotion and corresponds to past studies. There is a second force at work, however: subsequent tweets likely represent a late push as petitions approach the 100,000-signature parliamentary debate threshold.

This preliminary finding leads us to interrogate explicit one-mode network Twitter activity further. In particular, we look to tweet reception within individual users' follower networks as a way to see if their followers are engaged with the petitions. Both "favorites" and "retweets" have weak relationships with signatures; retweets with the stronger of the two (log-log correlation: 0.1527). This analysis is limited, however, as the tweets are collected relatively quickly after being posted to Twitter. Thus, there was limited (and slightly variable) time for other Twitter users to like or retweet a post before we observed it.

If there is a significant delay and tweet responses do not predict signatures, do the tweets simply refer to popular petitions after they have crossed the response threshold? Among all tweets, the median signatures of associated petitions is over 104,000 signatures: Twitter users overwhelmingly share popular petitions that have crossed the response threshold of 10,000 and indeed the Parliamentary debate threshold of 100,000. These thresholds are clearly affecting Twitter activity: as shown by Figure 7, the distribution of tweets is skewed towards those two thresholds. In particular, it appears tweets reference low-signature petitions and petitions that are crossing the 100,000 signature, Parliamentary debate, threshold. This finding complicates the above findings of a disconnect between Twitter activity and petitions: Twitter activity certainly responds to petition response thresholds. These findings will be unpacked in the Discussion section below.

### D. Users

Who tweets about petitions? What can we learn about political participation from these Twitter users? This section explores the some 250,000 Twitter users who used their accounts for explicit government interaction by sharing at least one petition in our sample. Figure 6 shows a word cloud of all user bios weighted for word occurrence. This permits us to draw some rough conclusions of, at the very least, how individuals describe themselves in their activity on Twitter. Football is a common theme: FC (football club), Football, LFC (Liverpool), and Fan are common descriptors. Also popular are emphases on family: family, wife, lover, and love are similarly common. Animal-lovers are similarly well represented, as discerned from: dog and animal. Finally, students are represented as well. This preliminary analysis indicates that users who tweet about petitions are not homogenous: they are not all young adults or male, for instance.

This variation leads us to question, what is driving these diverse people to participate politically on Twitter? This question is difficult to answer to available methods and data, but it leads us to a second question for explication: do users coalesce into discernable categories of interest based on their sharing activity? To find out, we run the Louvain Modularity community detection algorithm on our user network, where users are connected if they tweeted about the same petition, regardless of following/follower relationship. In contrast to the petition projection, we find robust clusters among users (modularity: 0.486/1.000).

Users divide into 239 clusters, 17 of which have over 1,000 users. See Figure 8 for relative distribution of users in all clusters. The largest cluster has over 93,000 users, and appears to heavily represent students and those from (or fans of the football club from) Liverpool.[11] Petitions addressing investigations into the Hillsborough disaster and efforts to save Grass Roots Football are likely responsible for this observed cluster. Significantly, however, we have no external benchmark for these findings: composition of clusters is compared to other petition-sharing clusters, not Twitter as a whole. The second largest cluster has over 34,000 users and the biographies of these users have numerous mentions of animals, wildlife, dogs, and cats. Other clusters are less clear in their composition; the third largest clusters with over 28,000 users is a good illustration of this (See Figure 9). In considering all clusters, some significant themes emerge. Some clusters represent self-identified wives and mothers while others represent fathers, raising the salience of gender in political Twitter participation. Important too are geographic identifiers: some clusters reference explicit locations like London, Liverpool, and Leicester, while others reference the entire United Kingdom. Other clusters have clear interests, including two religious clusters—one Christian and one Muslim (See Figures 10 and 11, respectively). There are similarly strong clusters among motor sport and cycling enthusiasts as well as those who enjoy the outdoors, fishing, and wildlife. Taken together, these findings indicate that users who share the same petition also tend to share self-described interests. Although users may not follow each other on Twitter, their expressed interests and political action align: in this way, Twitter petition sharing might be described in terms of an implicit interest group or latent support. This will be discussed further in the following section.

[11] These community analyses are qualitative in nature and draw on comparison of user descriptions. We do not seek to make quantitative claims in these descriptions.

This latent support characterizes the users network, but what of the users who are most integral to this network—who are they? Table 3 offers some details about the top 10 most central users in the user network of petition sharers. The most central user is a Twitter 'bot' that publicizes UK petitions: it tweeted 9437 times about 9084 unique petitions. The other most central users are average individuals—not bots nor organizations nor celebrities. One may expect that celebrities—verified accounts on Twitter—would be important sources of petition signatures. Yet, among the 1439 verified users in our sample, there is no relationship between celebrities tweeting about a petition and signatures (log correlation: 0.0043).[12] The most central individuals identify with politics in their bios: interests in politics, loyalty and duty to their country, and animal-rights causes are particularly notable. Yet these users are not using Twitter exclusively for political action: they identify their interests in culture, religion, family, music, and other things as well. These central users, aside from the bot, tended to tweet multiple times about petitions, as seen by comparing the right-most two columns in Table 3. Furthermore, their tweets were well received within their individual follower networks, as shown by the large number of retweets. As discussed below, these findings are quite important as they enrich the literature on those who share petitions.

[12] Verified users were also not central in the network, indicating that they tweeted about fewer unique petitions (log correlation: 0.0001).

## VI. Discussion

Taken together, our results indicate that Twitter serves to publicize petitions covering diverse topics, and such publicity is undertaken by a group of average individuals with varied political interests. The user network projection analyses addressed two lines of inquiry: (1) do commonly shared petitions reveal implicit interest groups of users and (2) who are the users more centrally implicated in Twitter-sharing activity? First, community detection yields high-modularity clusters of users that, when examining bios, reveals the clusters to represent themes in self-disclosed identity. Second, the most central users tended to be individuals with un-verified accounts and a range of—decidedly political—interests. Notably, the observed clusters of users is not necessarily explicit in Twitter following/follower relationships, but rather users who share the same petitions tend to share similar interests as expressed on their bios.

The petition network projection analyses addressed three primary questions: (1) are there emergent groups of petitions shared on the Twitter platform, (2) are commonly shared petitions likely to have similar signature counts, and (3) are petitions with many signatures on the UK petition website centrally shared among Twitter-sharing activity? First, community detection returns clusters of low modularity, indicating that there are not quality sub-clusters of petitions of any qualitative metric, content or otherwise. Second, we find a low correlation between edge weight (derived from common sharing users) and signature of connected petitions, meaning that users generally share petitions with both many and few signatures. Third, we find low correlation between petition centrality within the Twitter-sharing network and petition signatures on the UK website, but find a higher correlation between the two when limiting analyses to only petitions that crossed the first response threshold of 10,000 signatures. Taken together, these results indicate that petitions shared by the same user are not necessarily limited to a single topic, nor do they exclusively have high signature counts. The final result, which complicates discussion of responsivity to signature thresholds, is discussed below.

This study sought to understand Twitter as it relates to the UK petition platform. Results, in some respects reflect a larger tendency in unequal participation and activity on Twitter: despite a few users tweeting about many petitions, the median Twitter user tweeted about only one petition. These casual users help nuance our findings. For the median petition, there is a significant delay between when it is posted and when it is tweeted on Twitter. The underlying bimodal distribution adds nuance to the findings of Hale et al. (2013) and Yasseri et al. (2013). Although we do not analyze signatures over time in the current study, Figure 7 indicates that Twitter users are more responsive to the second, larger response threshold of 100,000 signatures for a parliamentary debate. Thus, the delay is not simply users tweeting about already successful petitions after the fact; instead, users are motivated by the potential for a parliamentary debate. This second threshold represents a more substantive benchmark than the primarily descriptive first threshold (See Bochel 2012) and a more explicit means of agenda-setting (Dumas et al. 2015); as such, it is unsurprising that users are more motivated at this threshold. This raises a question for further study: if average Twitter activity is motivated by this threshold, what about the most active users who tweet about many petitions—in what ways are they influenced by thresholds for different measures of success? Future study should address the role of different thresholds on different platforms and how these are impact the sharing of petitions on social media.

Indeed, active petition-sharing users constitute numerous latent support groups. Our user analyses reveal that such affiliations are implicit, as individuals do not band together in formal groups nor do they necessarily interact: rather, they both share the same petitions with their own follower networks. Making such common interests explicit and organizing around these already implicit interest groups could help drive explicit and purposive political mobilizations on Twitter in the future. In contrast to the typical notion of group formation proceeding collective action, we would see groups form through undertaking collection action. This is especially promising given that the most centrally located users in these petition-sharing networks are largely average individuals—not celebrities or formal organizations. This finding echoes findings of Strange (2011) who found that online petitions among global advocacy organizations served to foster substantive coordination between groups. Furthermore, our results differ from Lidner and Riehm (2011) who found that German online petitions exacerbate inequality of political participation; here, average individuals—both men and women—generally appear to be the most active. Indeed, such individuals could constitute a genuine and formidable political coalition if their relationships were to be made explicit. Future study should examine the extent to which these latent interest groups are, in fact, explicit: analyzing Twitter users' networks in relation to these two-mode analyses would be quite insightful.

This study is not without its limitations. Thirty-eight percent of successful petitions did not meet our Twitter significance threshold for inclusion in this study: these petitions were shared by other means. Google Analytics data from the UK Government Digital Service indicate that Twitter only accounted for approximately 20 percent of the referral traffic during this time period (Margetts et al., 2015). The construction of two-mode networks permits analysis of the interaction between micro- and macro- phenomena, but does so ex-post with an imposition of complete relationships. These relationships cannot be analyzed as they evolve over time in our work.

## VII. Conclusion

This work has sought to analyze sharing of government petitions via social media. We find that Twitter users who share UK government petitions share petitions with diverse topics. The most active users form implicit, latent interest groups based on their sharing activity and individual characteristics. These findings reveal the potential for substantive coordination and communication for political participation through social media. Such co-ordination would come after participation, rather than before, reversing the more structured and ordered vision of groups emerging from common interests and collective identity that characterized early democratic models of pluralism. Rather it would be something messier, with latent interest groups forming (and likely disbanding) in a more fluid, chaotic, and ad hoc manner, as they come together after undertaking 'tiny acts' of participation on social media —a model that Margetts et al. (2015) have called 'chaotic pluralism.'

Future studies should seek to expand our analyses to other social media platforms and countries and for wider time periods. Temporal analysis in our current study is also lacking: future analysis should examine the evolution of political participation on petition platforms over time. The 'trace data' from petitions, social media, and other digital platforms presents new opportunities to better understand the nature of politics.

### Table 1: Summary Statistics, by Petition

| | Signatures | Tweets | Users |
|---|---|---|---|
| **Median** | 10 | 1 | 1 |
| **Maximum** | 327,877 | 148,420 | 21,184 |

### Figure 1. Distribution of Signatures per Petition

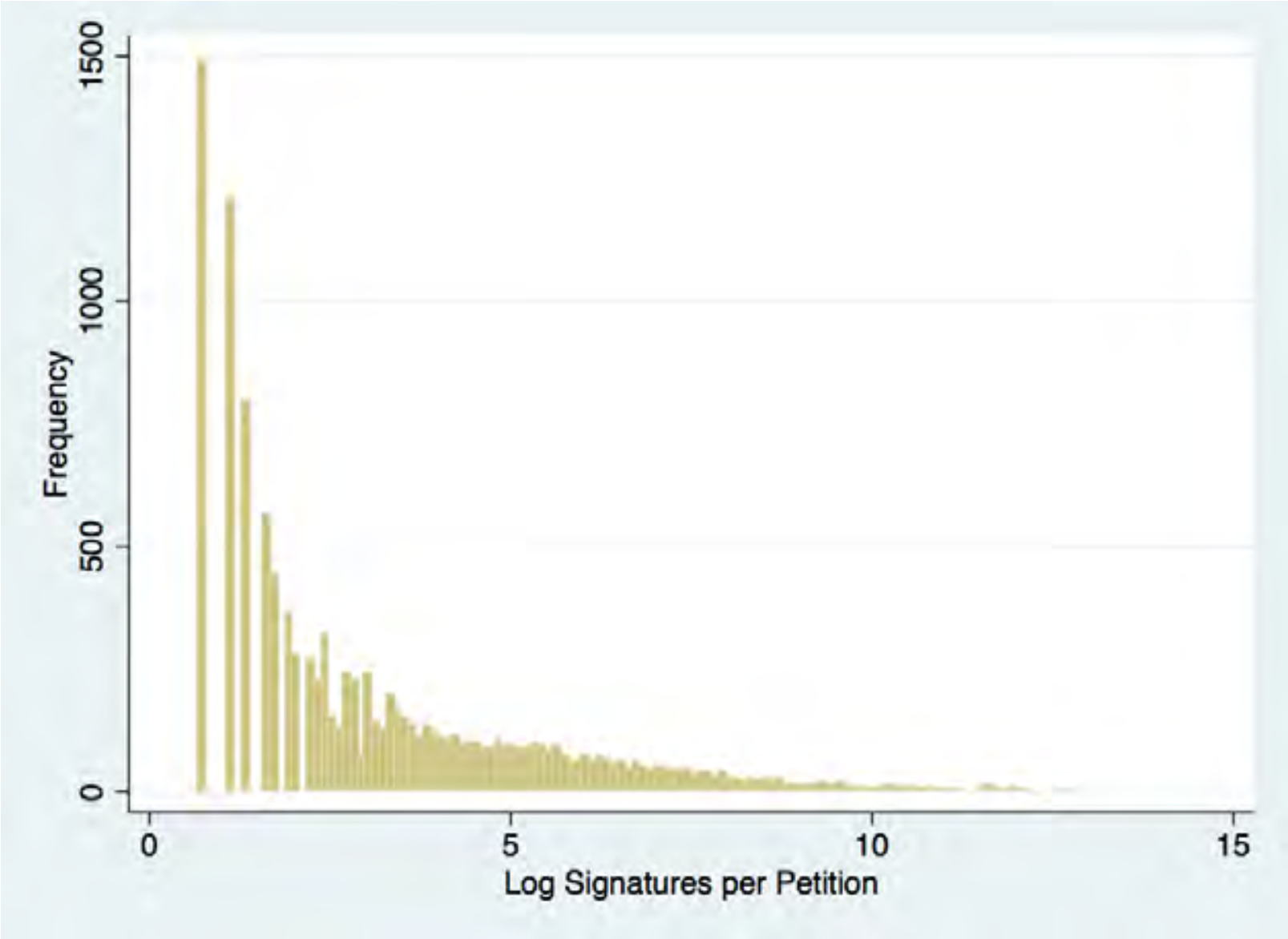

### Figure 2. Distribution of Tweets per Petition

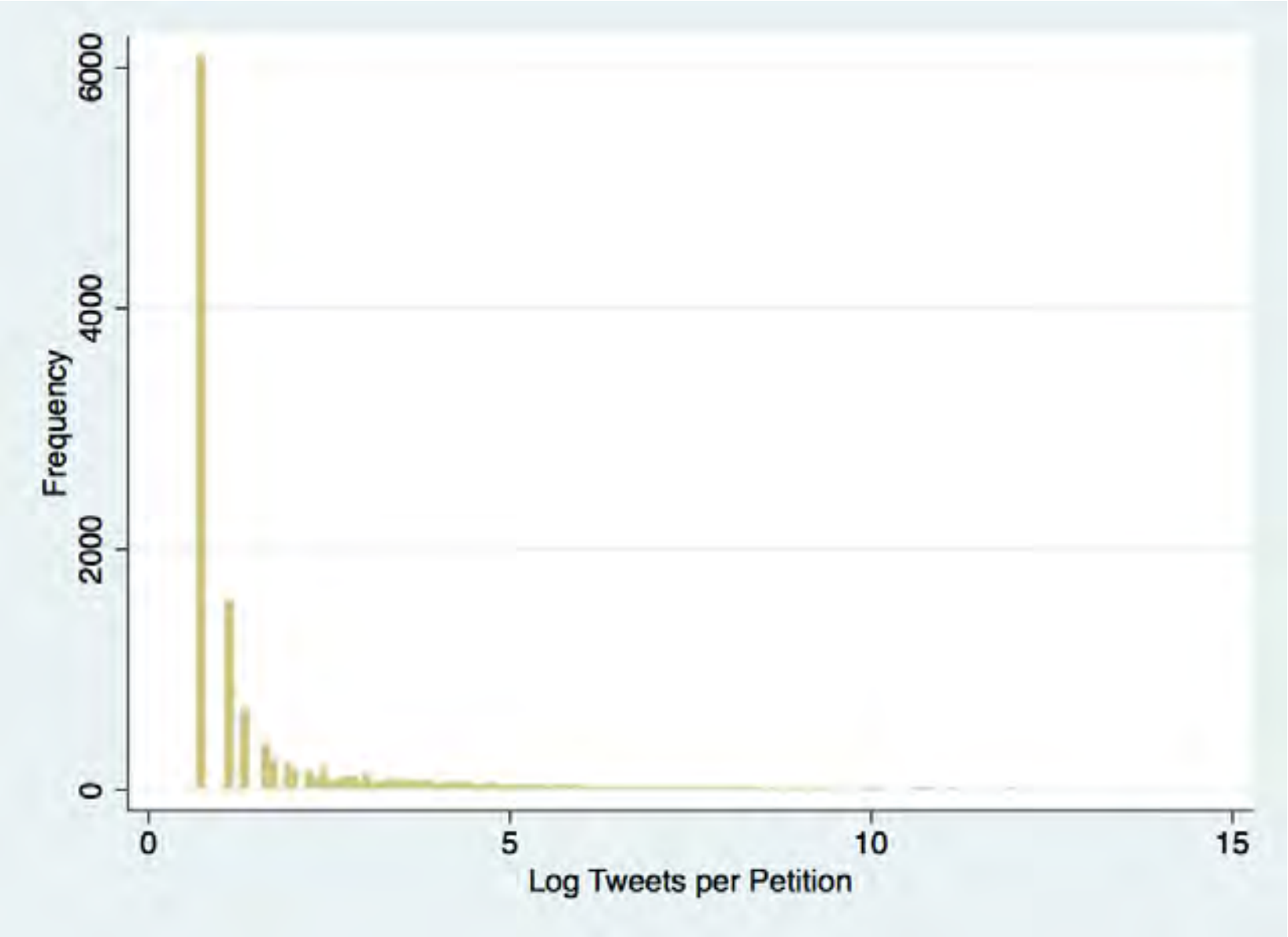

**Table 2: Regressions†**

| Predicting | Regr. 1 Signatures | Regr. 2 Signatures | Regr. 3 Signatures | Regr. 4 Signatures | Regr. 5 Signatures |
|---|---|---|---|---|---|
| Tweets | 1.144*** | -0.100** | | | -0.101** |
| | (0.0109) | (0.0461) | | | (0.0457) |
| Users | | 1.489*** | | | 1.474*** |
| | | (0.0536) | | | (0.0533) |
| Unique Audience | | | 0.481*** | | |
| | | | (0.00493) | | |
| Audience | | | | 0.462*** | |
| | | | | (0.00475) | |
| Constant | 1.368*** | 1.219*** | 0.667*** | 0.700*** | 3.636*** |
| | (0.0218) | (0.0217) | (0.0289) | (0.0287) | (0.503) |
| Department Fixed Effects | No | No | No | No | Yes |
| Observations | 11,405 | 11,405 | 11,405 | 11,405 | 11,405 |
| R-squared | 0.494 | 0.526 | 0.455 | 0.453 | 0.536 |

Standard errors in parentheses

*** p<0.01, ** p<0.05, * p<0.1

**† All variables are logarithm(variable)**

**Figure 3. Word Cloud of Titles of 178 Successful Petitions (Words Scaled for Occurrences)**

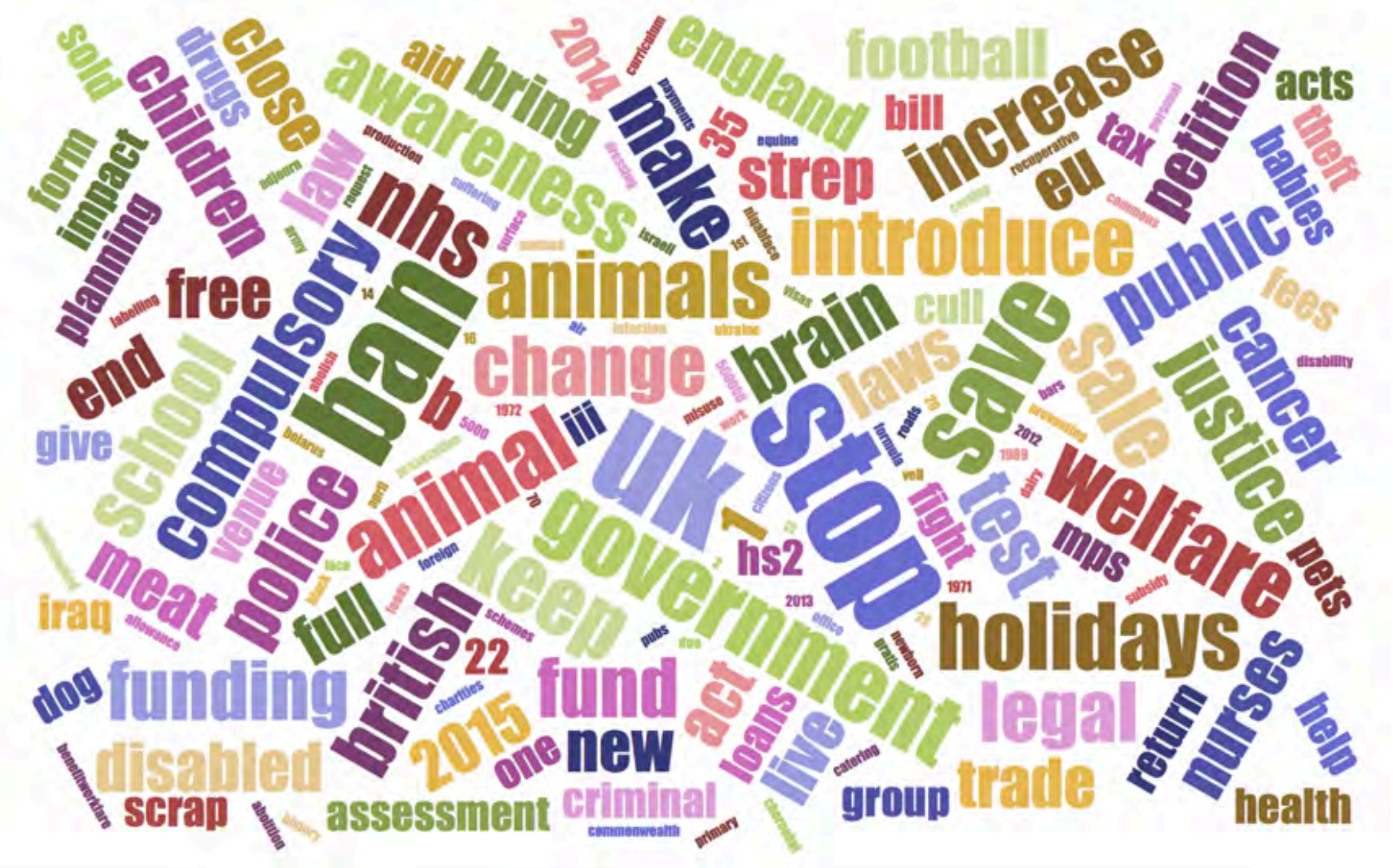

**Figure 4. PageRank of Petitions vs. Signatures**

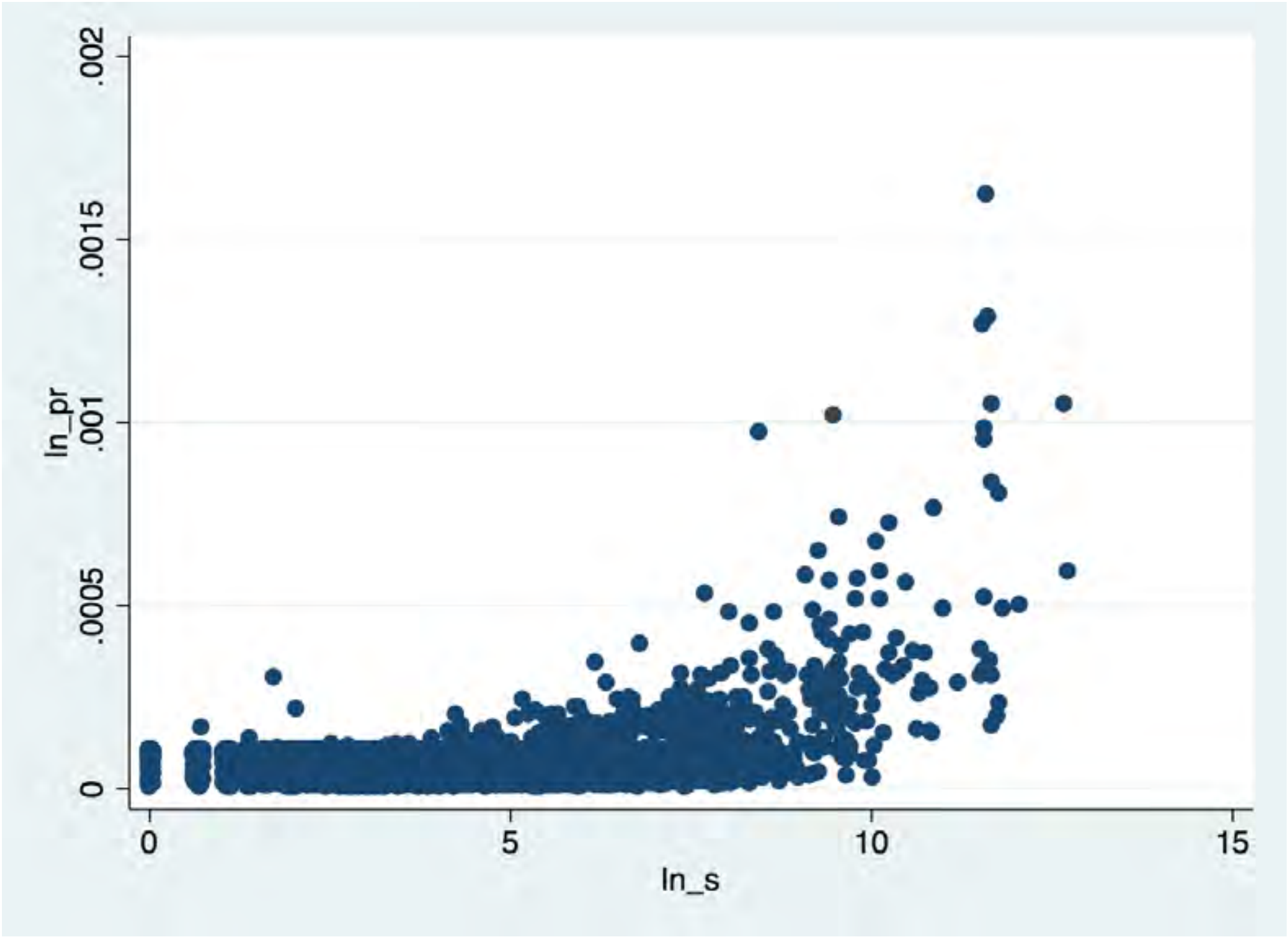

**Figure 5. Word Cloud of User Bios (Entire User Network)**

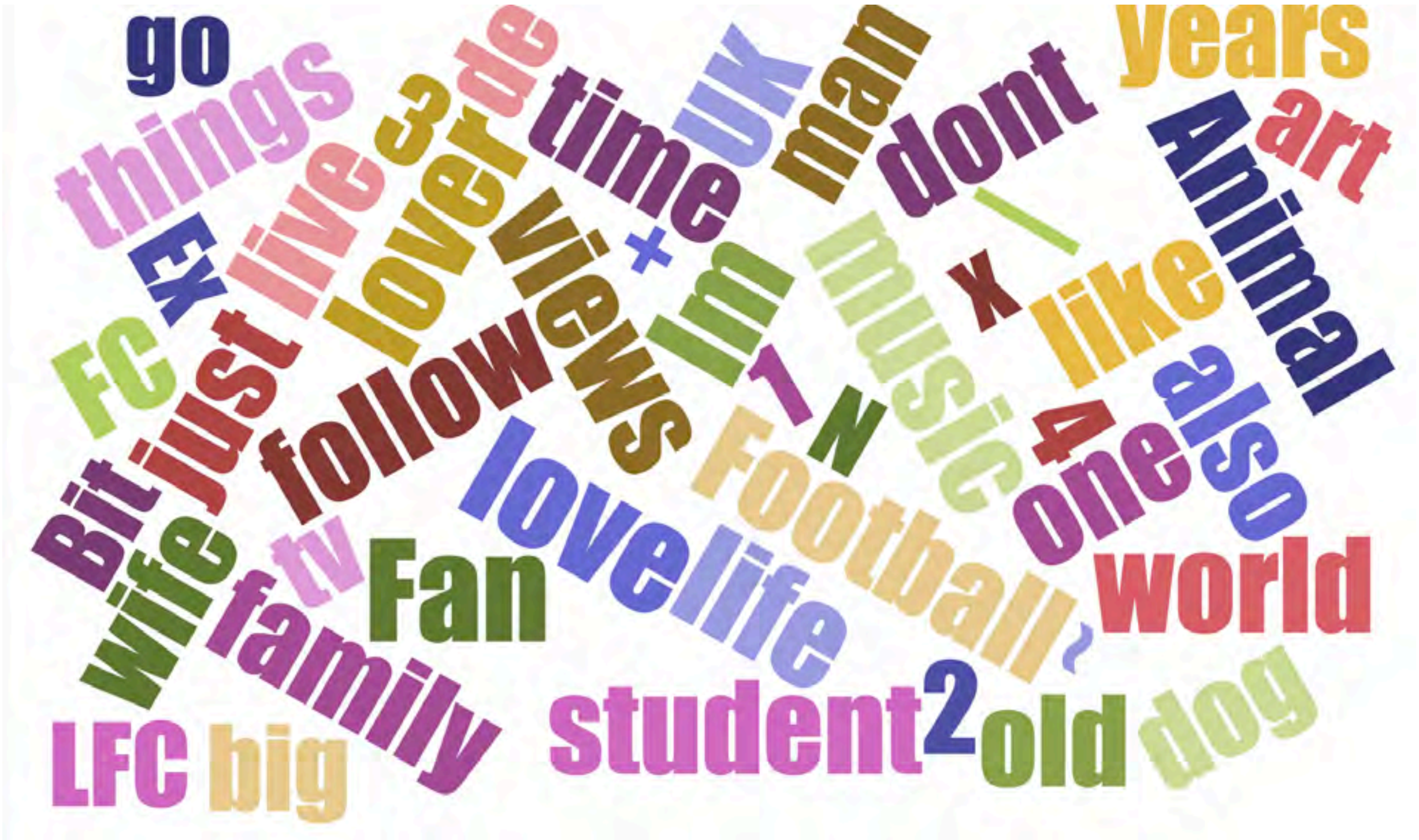

**Figure 6. Time Gap Between Petition Creation and Tweet (Excluding Outliers)**

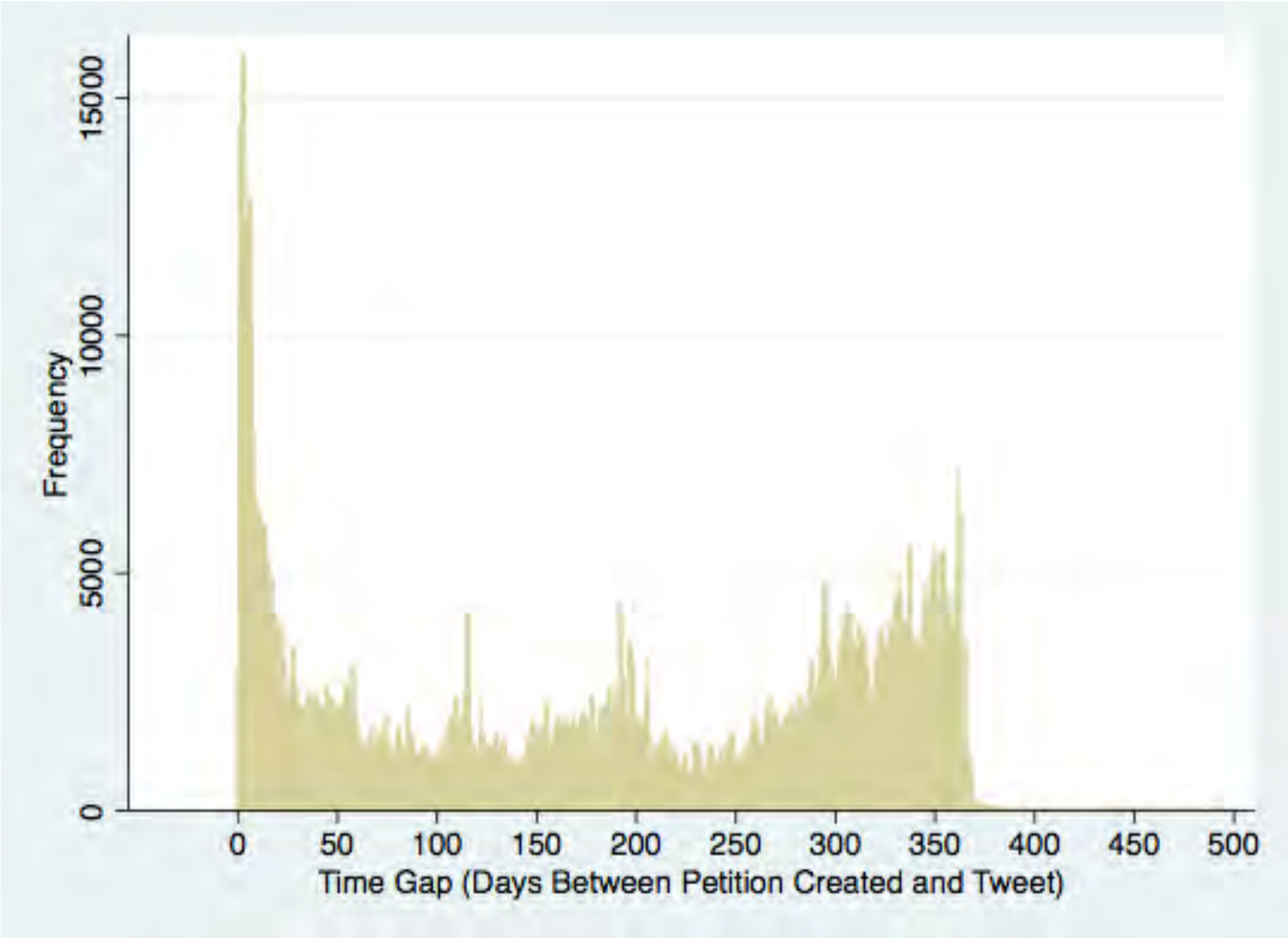

**Figure 7. Tweet Frequency of Petitions with Number of Signatures**

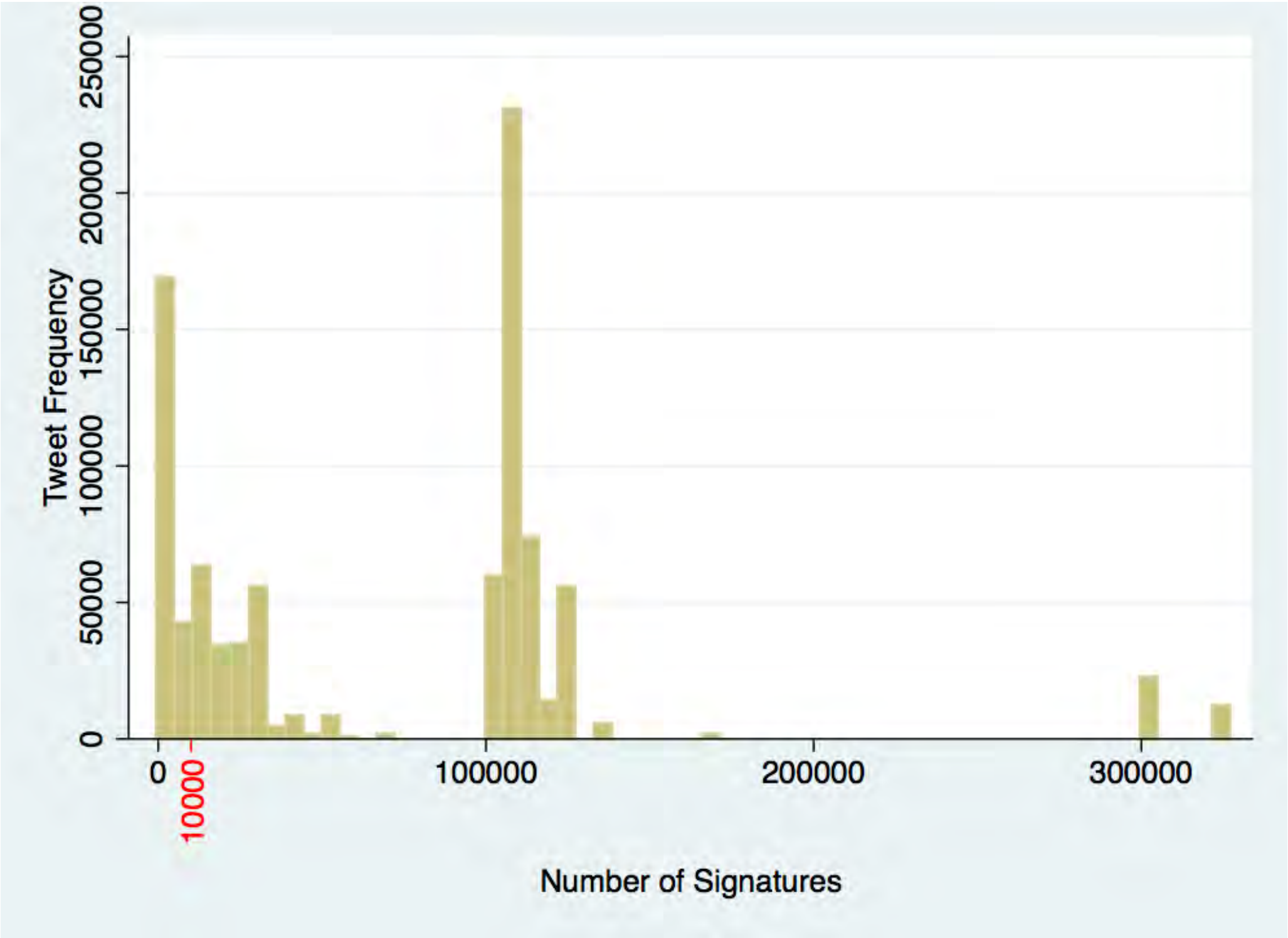

**Figure 8. Log Users per Cluster, Community Number on Horizontal Axis**

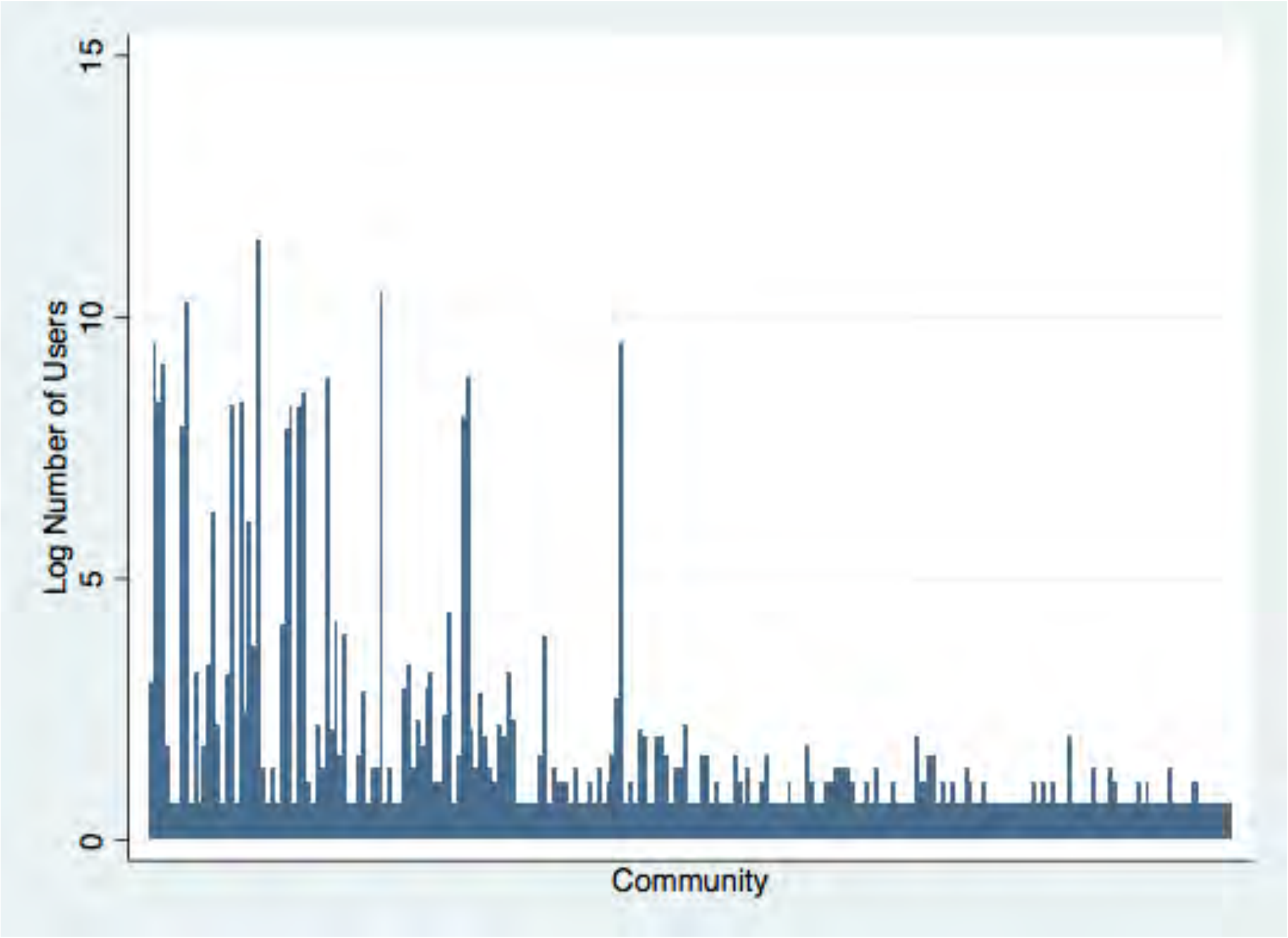

**Figure 9. Third Largest Cluster: 28,823 users' descriptions weighted for word occurrence**

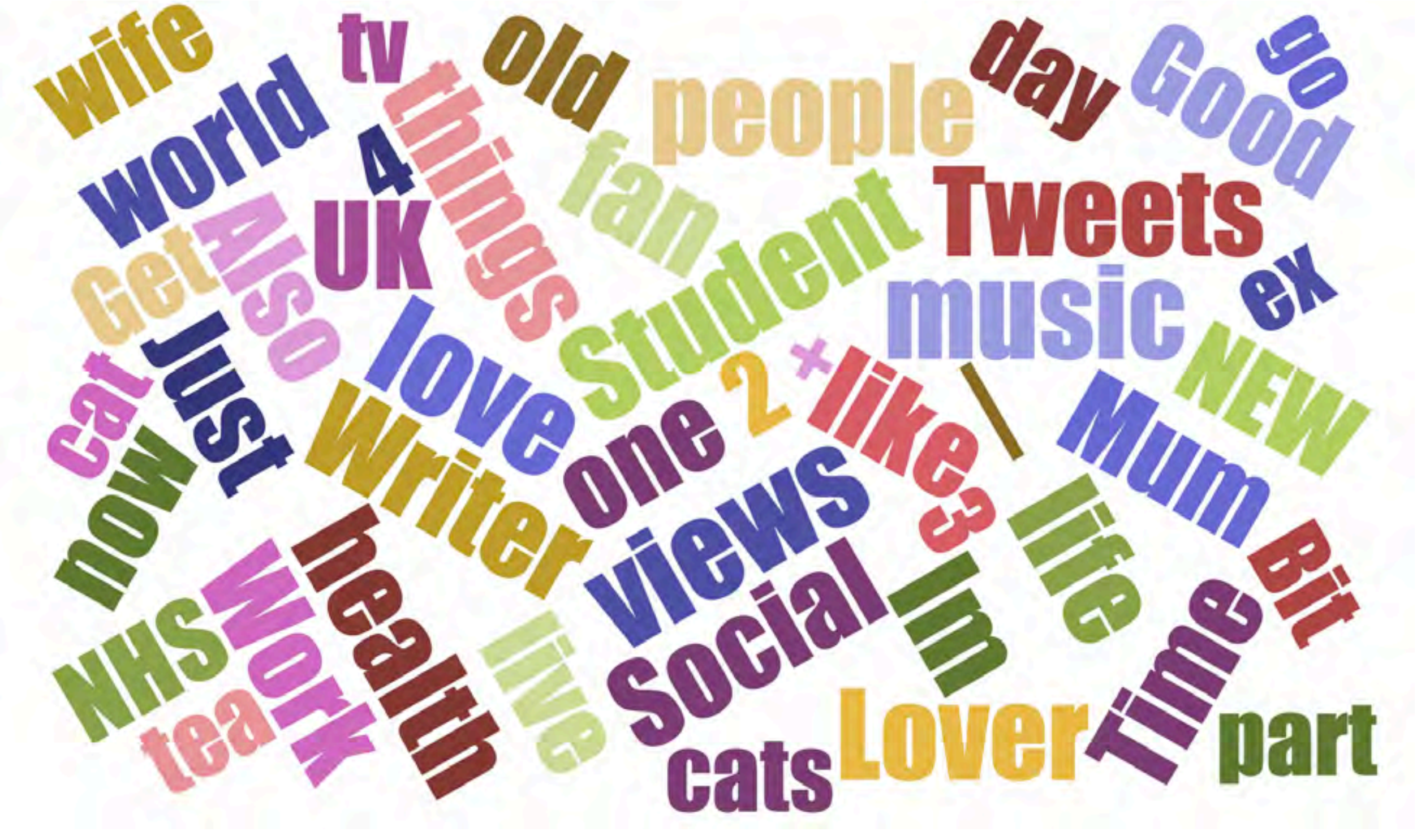

**Figure 10. Christian Users Cluster**

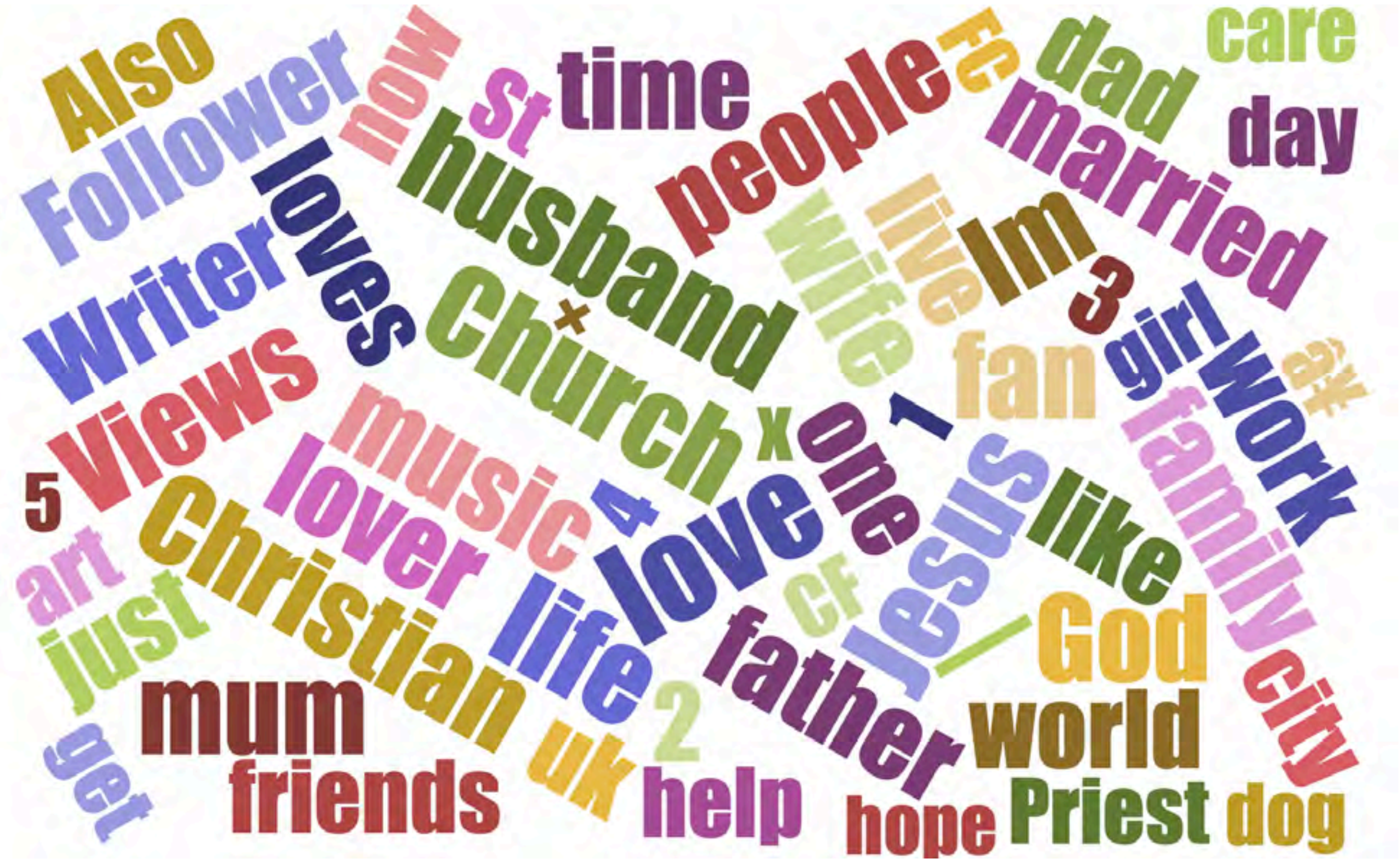

**Figure 11. Muslim Users Cluster**

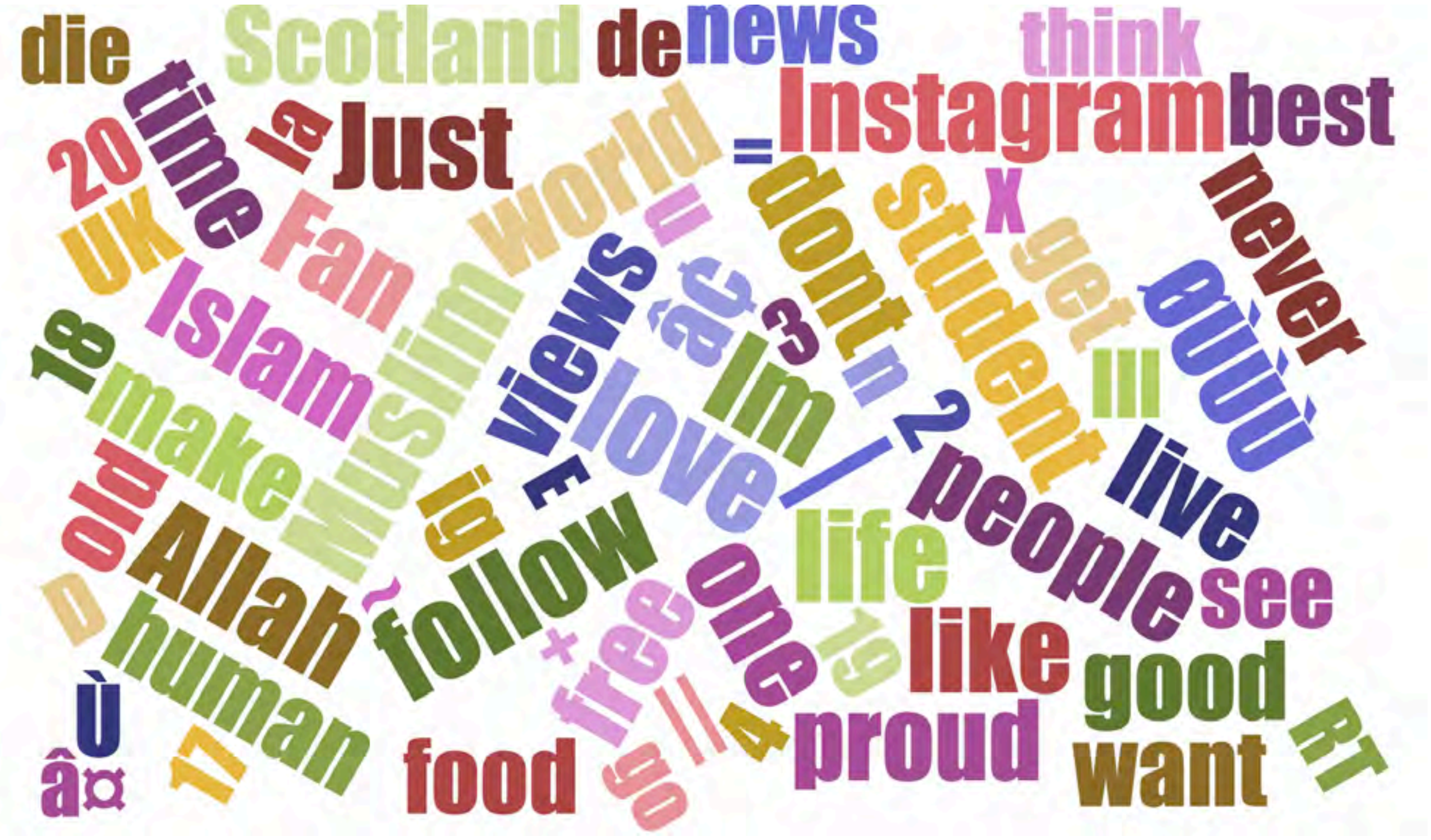

**Table 3. Profiles of the 10 Most Central Users in the User-Petition Sharing Network Projection**

| PageRank Centrality | User Bio | Followers | User Follows | Name | Petition Tweets Favorited by Followers | Petition Tweets Retweeted by Followers | Number of Tweets About Petitions | Number of Unique Petitions Tweeted |
|---|---|---|---|---|---|---|---|---|
| 0.0220134 | ePetitions scraped from http://t.co/0TdEJSqJnd | 0 | 12 | UK ePetitions | 43 | 38 | 9437 | 9084 |
| 0.0050271 | Help save England,England expects that everyman will do their duty,so lets not let her down.Love EnglandLove St Georges Day dont forget 23rd April.AFC Wimbledon | 1259 | 2001 | mpbaz9710 P.A.I | 0 | 1950 | 146 | 31 |
| 0.0036754 | Married to 'Peter Pan' 25 yrs in june, with 2 daughters my husband is 'still' the biggest kid the dog (Mollie) is just stupidly mard! Masive MU & westlife fan! | 977 | 2001 | Ingrid G. Price | 6 | 965 | 315 | 78 |
| 0.0025944 | Toriphile, Animal Activist, The Smiths, Neil Finn, Feminist, Cats, Burlesque, Gigs, Lush, NYR, Independent Cinema, Football, Musical Theatre, Forces Supporter. | 1360 | 1933 | PreciousThing | 0 | 4775 | 363 | 62 |
| 0.002436 | *(None provided)* | 71 | 148 | Barbara Langridge | 0 | 4889 | 367 | 49 |
| 0.0019731 | English Buddhist settled in Slovenia - happily married with grown up son - interested in the dharma and all good causes (human, animal, environmental). | 1683 | 1639 | Michael John Smith | 12 | 2511 | 351 | 46 |
| 0.0012313 | Determined to stop all cruelty to all living creatures as much as I possibly can | 1237 | 163 | caMORON's enemy | 6 | 5766 | 323 | 91 |
| 0.0011774 | Issues not political parties. Anti-corruption. Anti-bullying. Angry at Tory government. Unacceptable to stigmatise disabled & make their lives miserable. | 488 | 2001 | Hat Smith | 0 | 2981 | 187 | 56 |
| 0.0010484 | *(None provided)* | 14 | 110 | Debbie Harris | 0 | 8510 | 275 | 104 |
| 0.0009811 | Happily married with 2 grown-up sons. Love art and music and love painting and trying to play music. I am also interested in politics, comedy, disability. | 1823 | 1977 | Grace Murphy | 0 | 2939 | 103 | 47 |